\documentclass[reprint,floatfix,nofootinbib,amsmath,amssymb,unsortedaddress,superscriptaddress,aip,jcp]{revtex4-2}
\usepackage[utf8]{inputenc}
\usepackage{graphicx, cancel, booktabs, tabularx, amsmath, amsfonts, amssymb, bm, hyperref, placeins, float,mathrsfs}
\usepackage[normalem]{ulem}
\usepackage[table]{xcolor}
\usepackage{hyperref}

\newcommand{\op}{\hat{h}\left(\mathbf{x}\right)}
\newcommand{\paramOp}{\hat{h}^{\hat{\mathcal{P}}}\left(\alpha_1 \mathbf{x} + \beta_1\right)}
\newcommand{\phip}{\hat{\phi}^{\hat{\mathcal{P}}}\left(\mathbf{x}\right)}

\newcommand{\minval}[1]{\min\left(#1\right)}
\newcommand{\maxval}[1]{\max\left(#1\right)}
\begin{document}
\title{Recent advances in the SISSO method and their implementation in the SISSO++ code}
\author{Thomas A. R. Purcell}
\email{purcell@fhi-berlin.mpg.de}
\affiliation{The NOMAD Laboratory at the FHI of the Max-Planck-Gesellschaft and IRIS-Adlershof of the Humboldt-Universität zu Berlin, Faradayweg 4–6, D-14195 Berlin, Germany}

\author{Matthias Scheffler}
\affiliation{The NOMAD Laboratory at the FHI of the Max-Planck-Gesellschaft and IRIS-Adlershof of the Humboldt-Universität zu Berlin, Faradayweg 4–6, D-14195 Berlin, Germany}

\author{Luca M. Ghiringhelli}
\email{ghiringhelli@fhi-berlin.mpg.de}
\affiliation{Physics Department and IRIS-Adlershof, Humboldt Universit\"at zu Berlin, Zum Großen Windkanal 2 D-12489 Berlin, Germany} 
\affiliation{The NOMAD Laboratory at the FHI of the Max-Planck-Gesellschaft and IRIS-Adlershof of the Humboldt-Universität zu Berlin, Faradayweg 4–6, D-14195 Berlin, Germany}

\begin{abstract}
Accurate and explainable artificial-intelligence (AI) models are promising tools for the acceleration of the discovery of new materials, ore new applications for existing materials.
Recently, symbolic regression has become an increasingly popular tool for explainable AI because it yields models that are relatively simple analytical descriptions of target properties.
Due to its deterministic nature, the sure-independence screening and sparsifying operator (SISSO) method is a particularly promising approach for this application.
Here we describe the new advancements of the SISSO algorithm, as implemented into SISSO++, a C++ code with Python bindings.
We introduce a new representation of the mathematical expressions found by SISSO. 
This is a first step towards introducing ``grammar'' rules into the feature creation step.
Importantly, by introducing a controlled non-linear optimization to the feature creation step we expand the range of possible descriptors found by the methodology.
Finally, we introduce refinements to the solver algorithms for both regression and classification, that drastically increase the reliability and efficiency of SISSO.
For all of these improvements to the basic SISSO algorithm, we not only illustrate their potential impact, but also fully detail how they operate both mathematically and computationally.

\end{abstract}
\date{\today}
\maketitle

\maketitle

\section{Introduction}

Data-centric and artificial intelligence (AI) approaches are becoming a vital tool for describing physical and chemical properties and processes.
The key advantage of AI is its ability to find correlations between different sets of properties without the need to know which ones are important before the analysis.
Because of this, AI has become increasingly popular for materials discovery applications with uses in areas such as thermal transport properties~\cite{Zhu2021,Miller2017a}, catalysis~\cite{tran2018active,Foppa2021CatGEnes, Foppa2022CatRules}, phase stability~\cite{Bartel2019}, and quantum materials~\cite{Stanev2021}.
Despite the success of these methodologies, creating explainable and physically relevant AI models remains an open challenge in the field~\cite{Angelov2021,Gunning2019,das2020opportunities}.

One prevalent set of methods for explainable AI is symbolic regression~\cite{Xu2019,Aldeia2021,Holzinger2022,Li2021}.
Symbolic regression algorithms identify the optimal non-linear, analytic expressions for a given target property from a set of input features, i.e., the {\em primary features}, that are possibly related to the target~\cite{Wang-2019}.
Originally, (stochastic) genetic-programming-based approaches were and still are used to find these expressions~\cite{Koza-1994,Mueller-2014,Yuan-2017,Wang-2019}, but recently a more diverse set of solvers have been developed~\cite{Udrescu-2019,Kim2021SR,cranmer2019learning,valipour2021symbolicgpt,petersen2021deep,tenachi2023deep}.
The sure-independence screening and sparsifying operator (SISSO) approach combines symbolic regression with compressed sensing~\cite{Ouyang2019,Ouyang2017,Purcell2022,Foppa2022}, to provide a deterministic way of finding these analytic expressions.
This approach has been used to describe numerous properties including phase stability~\cite{Bartel2019,Ouyang2017,Schleder2020}, catalysis~\cite{Han2021}, and glass transition temperatures~\cite{Pilania2019}.
It has also been used in a multi-task~\cite{Ouyang2019} and hierarchical fashion~\cite{Foppa2022}.

The SISSO approach starts with a collection of primary features and mathematical unary and binary operators (e.g., addition, multiplication, $n$th root, logarithms, etc.). 
The first step is the {\em feature-creation} step, where a pool of {\em generated features}, is built by exhaustively applying the set of mathematical operators to the primary features. 
The algorithm is iteratively repeated by applying the set of operators to the primary features or the generated features created at the previous step. The number of iterations in this feature-creation step is called the rung.
The subsequent step is {\em descriptor identification}, i.e., compressed sensing is used to identify the best $n$-dimensional linear model by performing an $\ell_0$-regularized optimization on a subspace $\mathcal{S}$ of all generated expressions.
$\mathcal{S}$ is selected using sure-independence screening~\cite{Fan-2008}, with a suitable projection score, depending on whether one is solving a regression or classification problem (see below). 
For a regression problem this will rank all generated features according to their Pearson correlation values and select only the most correlated features, essentially performing a one-dimensional $\ell_0$-regularized linear regression on all features.
The result of the SISSO analysis is a $n-$dimensional descriptor, which is a vector with components from $\mathcal{S}$. 
For a regression problem, the SISSO model is the scalar product of the identified descriptor with the vector of linear coefficients resulting from the $\ell_0$-regularized linear regression. 
For a classification problem, the model is given as a set of hyperplanes that divide the points into classes, that are described by the scalar product of the identified descriptor, with a set of coefficients found by linear support vector machines (SVM).

Here, we introduce the new concepts implemented in the recently released SISSO++ code~\cite{Purcell2022} and detail their implementation.
Beyond creating a modular interface to run SISSO, SISSO++ also improves upon the algorithms in several aspects, for both the feature creation and descriptor identification steps.
The most important advancement of the code is expressing the features as binary expression trees, instead of strings, allowing us to recursively define all aspects of the generated expressions from the primary features.
With this implementation choice, we are able to create a complete description of the units for each generated feature, as well as an initial representation of its domain and range.
This allows for the creation of grammatically correct expressions, in terms of consistency of the physical units, and the control of numerical issues generated by features going out of their physically meaningful range.
In terms of the feature-creation step, we also discuss the implementation of {\em Parameteric SISSO}, which introduces the flexibility of non-linear parameters together with the operators that are optimized against a loss function based on the compressed-sensing-based feature selection metrics. 
This procedure was used to describe the thermal conductivity of a material in a recent publication~\cite{Purcell2022a}.
For the descriptor-identification step, we cover two components: an improved classification algorithm and the multi-residual approach.
For classification problems, we generalized the algorithm to work for any problem to an arbitrary dimension, and explicitly include a model identification via linear SVM.
The multi-residual approach, which was previously used in Ref.~\cite{Foppa2022}, introduces further flexibility for the identification of models with more than one dimension. 
Here, we provide a in-depth discussion of its machinery.

\section{Feature Creation}
\label{sec:fc}

\subsection{Binary-Expression-Tree Representation of Features}
\label{subsec:fc_bet}

\begin{figure}
    \centering
    \includegraphics{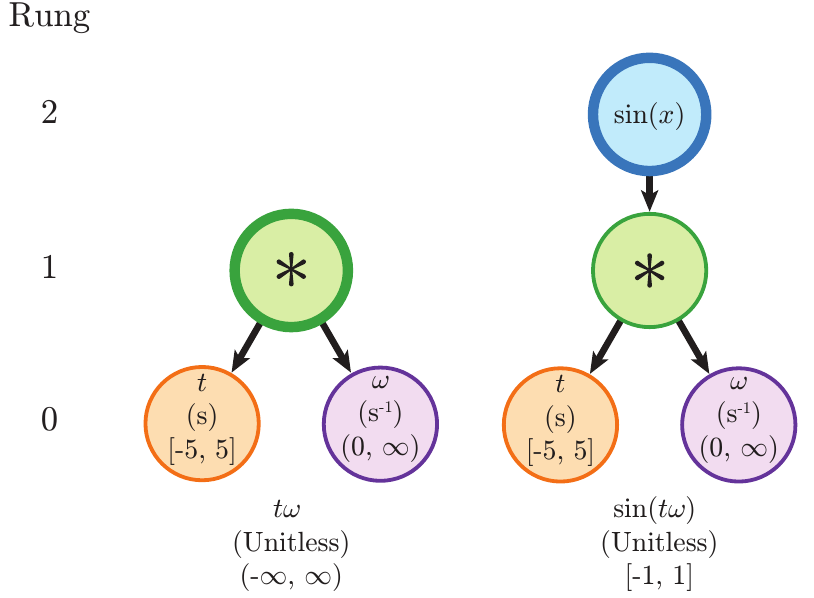}
    \caption{A demonstration of the new representation of the features in the SISSO++ code. The feature is stored as the root of the tree (represented by the thick border), the primary features are the leaves, and the rung corresponds to the height of the tree, i.e. the longest path between each leaf and the root. The unit, range, expressions, and values of the features are necessarily stored only for the primary features, with them defined recursively for all generated expressions.}
    \label{fig:feature_demonstration}
\end{figure}

The biggest advancement of the implementation of SISSO in SISSO++, compared to the original implementation in Ref. \cite{Ouyang2017}, is its modified representation of the features as binary expression trees, instead of strings.
This representation is illustrated in Figure~\ref{fig:feature_demonstration} and easily allows for all aspects of the generated features to be recursively calculated on the fly from the data stored in the primary features.
For certain applications, it is also possible to store the data of higher-rung features, to reduce the overall computational cost of the calculations.
The individual features are addressed by the root node of the binary expression tree, and stored in the code as a \textsc{shared\_ptr} from the C++ standard library.
This representation reduces the overall memory footprint of each calculation, as the individual features only need to be created once and only copies of shared pointers need to be stored for each new expression.
The remainder of this section will be used to describe the various aspects of the new representation including a description of the units and range of the features, as well as how it is used to generate the feature space.

\subsubsection{Units}
\label{subsec:fc_units}

An important upgrade in SISSO++ is its generalized and exact treatment of units for the expressions.
In physics, dimensional analysis is an important tool when generating physically meaningful expressions, and it is necessary to include it when using symbolic regression for scientific problems.
We introduced this into SISSO++ by determining the units for each new expression from the primary features, and explicitly checking to ensure that a new expression is physically possible.
Within the code the \textsc{Units} are implemented as dictionaries with the key representing the base unit and the value representing the exponent for each key, e.g., $\mathrm{m/s^2}$ would be stored as $\mathrm{\{m: 1, s: -2\}}$.
Functions exist to transform the \textsc{Units} to and from strings to more easily represent the information.
We then implemented a multiplication, division, and power functions for these specialized dictionaries, allowing for the units of the generated features to be derived recursively following the rules in Table~\ref{tab:unit_construction}.
An important caveat is that the current implementation  can not convert between two units for the same physical quantity units, e.g. between nanometers, picometers, and Bohr radii.

\begin{table}
\centering
\caption{How the units are calculated for each operation}
\label{tab:unit_construction}
\begin{tabular}{p{2.0cm} l}
\toprule
Operation            & Resulting Unit    \\
\midrule
$A + B$              & Unit(A)           \\
$A - B$              & Unit(A)           \\
$A \left(B\right)$   & Unit(A) * Unit(B) \\
$A / B$              & Unit(A) / Unit(B) \\
$\left|A - B\right|$ & Unit(A)           \\
$\left|A\right|$     & Unit(A)           \\
$\sin\left(A\right)$ & Unitless          \\
$\cos\left(A\right)$ & Unitless          \\
$\exp\left(A\right)$ & Unitless          \\
$\exp\left(-A\right)$& Unitless          \\
$\log\left(A\right)$ & Unitless          \\
$\left(A\right)^{-1}$& Unit(A)$^{-1}$    \\
$\left(A\right)^2$   & Unit(A)$^{2}$     \\
$\left(A\right)^3$   & Unit(A)$^{3}$     \\
$\left(A\right)^6$   & Unit(A)$^{6}$     \\
$\sqrt{A}$           & Unit(A)$^{1/2}$   \\
$\sqrt[3]{A}$        & Unit(A)$^{1/3}$   \\
\botrule
\end{tabular}
\end{table}

Using this implementation of units, a minimal treatment of dimensional analysis can be performed in the code.
The dimensional analysis focuses on whether the units are consistent within each expression and for the final model.
For the final model, this check is used to determine the units of the fitted constants in the linear models at the end, which can take arbitrary units, and therefore can do any unit conversion natively.
The restrictions, used to reject expressions by dimensional analysis, are outlined in Table~\ref{tab:unit_restrictions}, and can be summarized as only allowing addition and subtraction to act on features of the same units, and all transcendental operations must act on a unitless quantity.
With these two restrictions in place, only physically possible features can be found, and the choice of units should no longer affect which features are selected.
If one wants to allow for non-physical expressions to be found, this can be achieved by providing all input primary features with no units associated to them.

\begin{table}
\centering
\caption{Restrictions for each unit, if an operation is not listed there are no restrictions.}
\label{tab:unit_restrictions}
\begin{tabular}{p{2.0cm} l}
\toprule
Operation            & Unit Restriction       \\
\midrule
$A + B$              & Unit(A) == Unit(B)     \\
$A - B$              & Unit(A) == Unit(B)     \\
$\left|A - B\right|$ & Unit(A) == Unit(B)     \\
$\sin\left(A\right)$ & Unit(A) == $\emptyset$ \\
$\cos\left(A\right)$ & Unit(A) == $\emptyset$ \\
$\exp\left(A\right)$ & Unit(A) == $\emptyset$ \\
$\exp\left(-A\right)$& Unit(A) == $\emptyset$ \\
$\log\left(A\right)$ & Unit(A) == $\emptyset$ \\
\botrule
\end{tabular}
\end{table}

\subsubsection{Range}
\label{subsec:fc_range}

Another important advancement of the feature-creation step is the introduction of ranges for the primary features, which act as a domain for future operations during feature creation.
One of the challenges associated with symbolic regression, especially with smaller datasets, is that the selected expressions can sometimes contain discontinuities that are outside of the training data, but still within the relevant input space for a given problem.
For example, this can lead to an expression taking the logarithm of a negative number, resulting in an undefined prediction.
SISSO++ solves this problem by including an option for describing the range of a primary feature in standard mathematical notation, e.g. $\left[\left. 0, \infty\right)\right.$, and then using that to calculate the range for all generated formulas using that primary feature, following the rule specified in Table~\ref{tab:range_definitions}.
In the code, the ranges are referenced as the \textsc{Domain} because the range for a feature of rung $n-1$ is the domain for a possible expression of rung $n$ that is using that feature.
While all ranges in Table~\ref{tab:range_definitions} assume inclusive endpoints, the implementation can handle both exclusive endpoints and a list of values explicitly excluded from the range, e.g., point discontinuities inside the primary features themselves.

\begin{table*}
\centering
\caption{How the range for each operation is calculated}
\label{tab:range_definitions}
\begin{tabular}{p{2.0cm} l}
\toprule
Operation            & Resulting Range    \\
\midrule
$A + B$              &  $\left[ \minval{A} + \minval{B}, \maxval{A} + \maxval{B} \right] $ \\
$A - B$              &  $\left[ \minval{A} - \maxval{B}, \maxval{A} - \minval{B} \right] $ \\
$A \left(B\right)$   & $\left[\minval{\minval{A} * \minval{B}, \minval{A} * \maxval{B}, \maxval{A} * \minval{B}, \maxval{A} 
                        * \maxval{B}}, \right.$ \\
                     & $\left.\maxval{\minval{A} * \minval{B}, \minval{A} * \maxval{B}, \maxval{A} * \minval{B}, \maxval{A} 
                        * \maxval{B}} \right]$ \\
$A / B$              &  $\left[\mathrm{Range}\left(A\right) * \mathrm{Range}\left(B^{-1}\right)\right]$ \\
$\sin\left(A\right)$ &  $\left[-1, 1\right]$ \\
$\cos\left(A\right)$ &  $\left[-1, 1\right]$ \\
$\exp\left(A\right)$ &  $\left[\exp\left(\minval{A}\right), \exp\left( \maxval{A} \right) \right]$ \\
$\exp\left(-A\right)$&  $\left[\exp\left(-\maxval{A}\right), \exp\left(-\minval{A} \right) \right]$ \\
$\log\left(A\right)$ &  $\left[\log\left(\minval{A}\right), \log\left( \maxval{A} \right) \right]$ \\
$\left(A\right)^{-1}$& if($0 \in \mathrm{Range}\left(A\right)$): \\
                     & ~~~~ $\left(-\infty, 0\right) \cup \left(0, \infty\right)$ \\
                     & else: \\
                     & ~~~~ $\left[\left(\maxval{A}\right)^{-1}, \left( \minval{A} \right)^{-1} \right]$ \\
$\left(A\right)^3$   &  $\left[\left(\minval{A}\right)^3, \left( \maxval{A} \right)^3 \right]$ \\
$\sqrt{A}$           &  $\left[\sqrt{\minval{A}}, \sqrt{\maxval{A}} \right]$ \\
$\sqrt[3]{A}$        &  $\left[\sqrt[3]{\minval{A}}, \sqrt[3]{\maxval{A}} \right]$ \\
$\left|A\right|$     &  $\left[\maxval{0, \minval{A}}, \maxval{\left|\maxval{A}\right|, \left|\minval{A}\right|} \right]$\\
$\left(A\right)^2$   &  $\left[\maxval{0, \minval{A}}^2, \maxval{\left|\maxval{A}\right|, \left|\minval{A}\right|}^2 \right]$ \\
$\left(A\right)^6$   &  $\left[\maxval{0, \minval{A}}^6, \maxval{\left|\maxval{A}\right|, \left|\minval{A}\right|}^6 \right]$\\
$\left|A - B\right|$ & $\left[\maxval{0, \minval{A} - \maxval{B}}, \right.$ \\
                     & $\left.\maxval{\left|\maxval{\maxval{A} - \minval{B}}\right|, \left|\minval{\maxval{A} - \minval{B}}\right|} \right]$ \\

\botrule
\end{tabular}
\end{table*}

Table~\ref{tab:range_restrictions} lists the cases where the range of a feature is used to prevent a new expression from being generated.
In all cases, this prevents an operation from occurring where a mathematical operation would be not defined, such as taking the square-root of a negative number.
In cases where the range of values for a primary feature is not defined, then these checks are not performed and the original assumption that all operations are safe is used.

\begin{table}
\centering
\caption{Domain restrictions for each operation, if an operation is not listed there are no restrictions.}
\label{tab:range_restrictions}
\begin{tabular}{p{2.0cm} l}
\toprule
Operation             & Domain Restriction       \\
\midrule
$A / B$               & $0 \notin \mathrm{Range\left(B\right)}$         \\
$\left(A\right)^{-1}$ & $0 \notin \mathrm{Range\left(A\right)}$         \\
$\log\left(A\right)$  & $\minval{\mathrm{Range}\left(A\right)} > 0 $    \\
$\sqrt{A}$            & $\minval{\mathrm{Range}\left(A\right)} \geq 0 $ \\
\botrule
\end{tabular}
\end{table}

\subsection{Parametric SISSO}
\label{subsec:fc_param}
\begin{figure}
    \centering
    \includegraphics{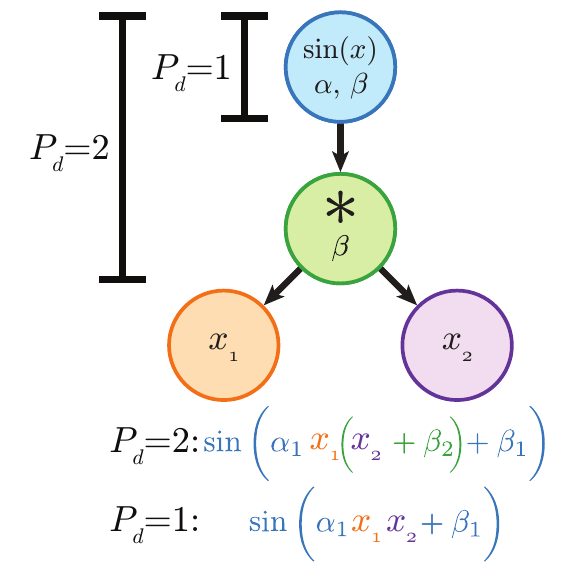}
    \caption{A graphical representation of the effect of the parameterization depth. If $P_d=1$ then only the $\mathrm{sin}$ operator gets parameterized, while if $P_d=2$ both operations are parameterized}
    \label{fig:si_param_depth_def}
\end{figure}
Parametric SISSO extends the feature creation step of SISSO to automatically include scale and bias terms for each operation, as used in Purcell {\em et al.}\cite{Purcell2022a}.
For a general operator, $\op \in \hat{\mathcal{H}}$, with a set of scale and bias parameters, $\hat{\mathcal{P}}$, the parameterization scheme updates the operator to be,
\begin{equation}
    \op \to \paramOp,
\end{equation}
where $\alpha_1$ is the scale parameter, $\beta_1$ is the bias term, and $\mathbf{x}$ is a vector containing all input data.
These new operators can then be used to create a new feature, $\phip$, as is normally done in SISSO, where each feature has its own set of parameters.
However, this does introduce a new hyperparameter for the feature creation step, the parameterization depth, $P_d$, which defines the level at which  the parameterization occurs in the binary expression tree.
This is best described in Figure~\ref{fig:si_param_depth_def}, where $P_d$ controls which operations are included in the parameterization.
In this example, if $\beta_2$ was previously set by another optimization and $P_d=2$, then that previous value will be ignored for this feature only; however, if $P_d=1$, then the existing $\beta_2$ will be preserved.
In order to avoid linear dependencies between different operations and the constants set during linear regression, some of the scale and bias terms are set to one and zero, respectively, for a summary of this for all operators see Table~\ref{tab:free_parameters}.
It is important to note that for the log operator, $\alpha$ is always set to $\pm1$ to avoid linear dependencies with other parameters.
Although this does leave a unit dependency, it can be removed with
\begin{equation}
    \ln\left(x + \beta\right) \rightarrow \ln\left(\alpha_\mathrm{unit} \left(x + \beta\right)\right) - \ln\left(\alpha_\mathrm{unit}\right),
\end{equation}
where $\alpha_\mathrm{unit}$ is the unit conversion factor.

\begin{figure*}
    \centering
    \includegraphics{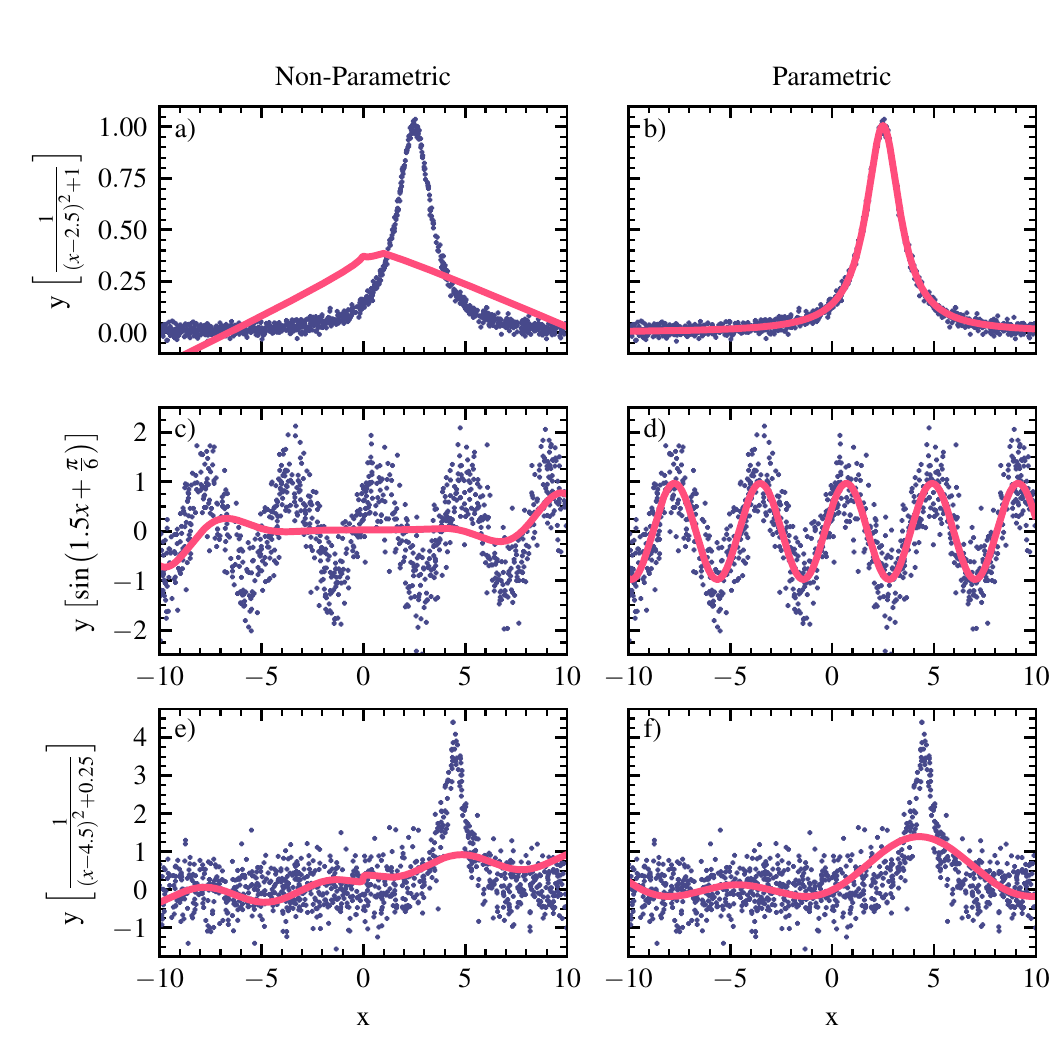}
    \caption{A comparison of the expressions found non-parametric (a, c, and e) and parametric SISSO (b, d, and f) for a Lorentzian (a, b, e, and f) and sin (c, d) function. Blue dots represent the training data, and the red line represents the expressions found by SISSO. The parameterization scheme either finds the correct (b and d) or better model (f) than the non-parametric functions, even when  the high noise or bad initial guess of the parameters leads to a non-optimized solution.}
    \label{fig:param_results}
\end{figure*}

\begin{table}
\centering
\caption{The free parameters and updated equations for each operation used in the updated feature creation step of SISSO}
\label{tab:free_parameters}
\begin{tabular}{p{2.0cm} l}
\toprule
Operation            & Parameterized Operation                           \\
\midrule
$A + B$              & $A + \alpha B$                                    \\
$A - B$              & $A - \alpha B$                                    \\
$A \left(B\right)$              & $A\left(B + \beta\right)$                      \\
$A / B$              & $A / \left(B + \beta\right)$                      \\
$\left|A - B\right|$ & $\left|A - \left(\alpha B + \beta\right)\right|$  \\
$\left|A\right|$     & $\left|A + \beta\right|$ \\
$\sin\left(A\right)$ & $\sin\left(\alpha A + \beta\right)$               \\
$\cos\left(A\right)$ & $\cos\left(\alpha A + \beta\right)$               \\
$\exp\left(A\right)$ & $\exp\left(\alpha A\right)$\footnote[1]{$\alpha > 0$} \\
$\exp\left(-A\right)$& $\exp\left(-\alpha A \right)$\footnotemark[1] \\
$\log\left(A\right)$ & $\log\left(A + \beta\right)$\footnote[2]{$\alpha$ can only be 1.0 or -1.0} \\
$\left(A\right)^{-1}$& $\left(A + \beta\right)^{-1}$                     \\
$\left(A\right)^2$   & $\left(A + \beta\right)^2$                        \\
$\left(A\right)^3$   & $\left(A + \beta\right)^3$                        \\
$\left(A\right)^6$   & $\left(A + \beta\right)^6$                        \\
$\sqrt{A}$          & $\sqrt{\alpha A + \beta}$\footnotemark[2]              \\
$\sqrt[3]{A}$       & $\sqrt[3]{A + \beta}$                              \\
\botrule
\end{tabular}
\end{table}

Once $\phip$ is defined, all parameters $\hat{p}\in\hat{\mathcal{P}}$ are optimized using the non-linear optimization library NLopt~\cite{NLopt}.
We use the Cauchy loss function as the objective for the optimization 
\begin{subequations}
\begin{align}
    \underset{\hat{\mathcal{P}}}{\mathrm{min}} & \,f\left(\mathbf{P}, \hat{\phi}^{\hat{\mathcal{P}}}\right)\\
    f\left(\mathbf{P}, \hat{\phi}^{\hat{\mathcal{P}}}\right) & =\sum_i^{n_\mathrm{samp}} \frac{c^2}{n_\mathrm{samp}} \log\left(1 + \left(\frac{P_i - \hat{\phi}^{\hat{\mathcal{P}}}\left(\mathbf{x}_i\right)}{c}\right)^2\right),
    \label{objective}
\end{align}
\label{eq:objective}
\end{subequations}
where $\mathbf{P}$ is a property vector, $c$ is a scaling factor set to 0.5 for all calculations and $n_\mathrm{samp}$ is the number of samples.
We chose to use the Cauchy loss function over the mean square error to make the non-linear optimization more robust against outliers in the dataset.
Because Equation~\ref{eq:objective}b is not scale or bias invariant, additional external parameters $\alpha_{ext}$ and $\beta_{ext}$ are introduced to respectively account for these effects.
For the case of multi-task SISSO~\cite{Ouyang2019}, each task has its own external bias and scale parameters to account for the individual linear regression solutions.
As an example for the feature illustrated in Figure~\ref{fig:si_param_depth_def} ($P_d=2$), the function that is optimized would be
\begin{equation}
    \hat{\phi}^{\hat{\mathcal{P}}}\left(\mathbf{x}\right) = \alpha_{ext} \sin\left(\alpha_1 \frac{x_1}{x_2 + \beta_2} + \beta_{1}\right) + \beta_{ext}.
\end{equation} 
To initialize the parameters in $\hat{\mathcal{P}}$, we set all internal $\alpha$ and $\beta$ terms are set to 1.0 and 0.0, respectively, and $\alpha_{ext}$ and $\beta_{ext}$ are set to the solution of the least squares regression problem for each task.
In some cases, $\beta$ can be set to a non-zero value if leaving it at zero would include values outside the Domain of the operator. 
In these cases, $\beta$ is set to $\minval{\mathbf{x}} + 10^{-10}$.
\begin{figure*}
    \centering
    \includegraphics{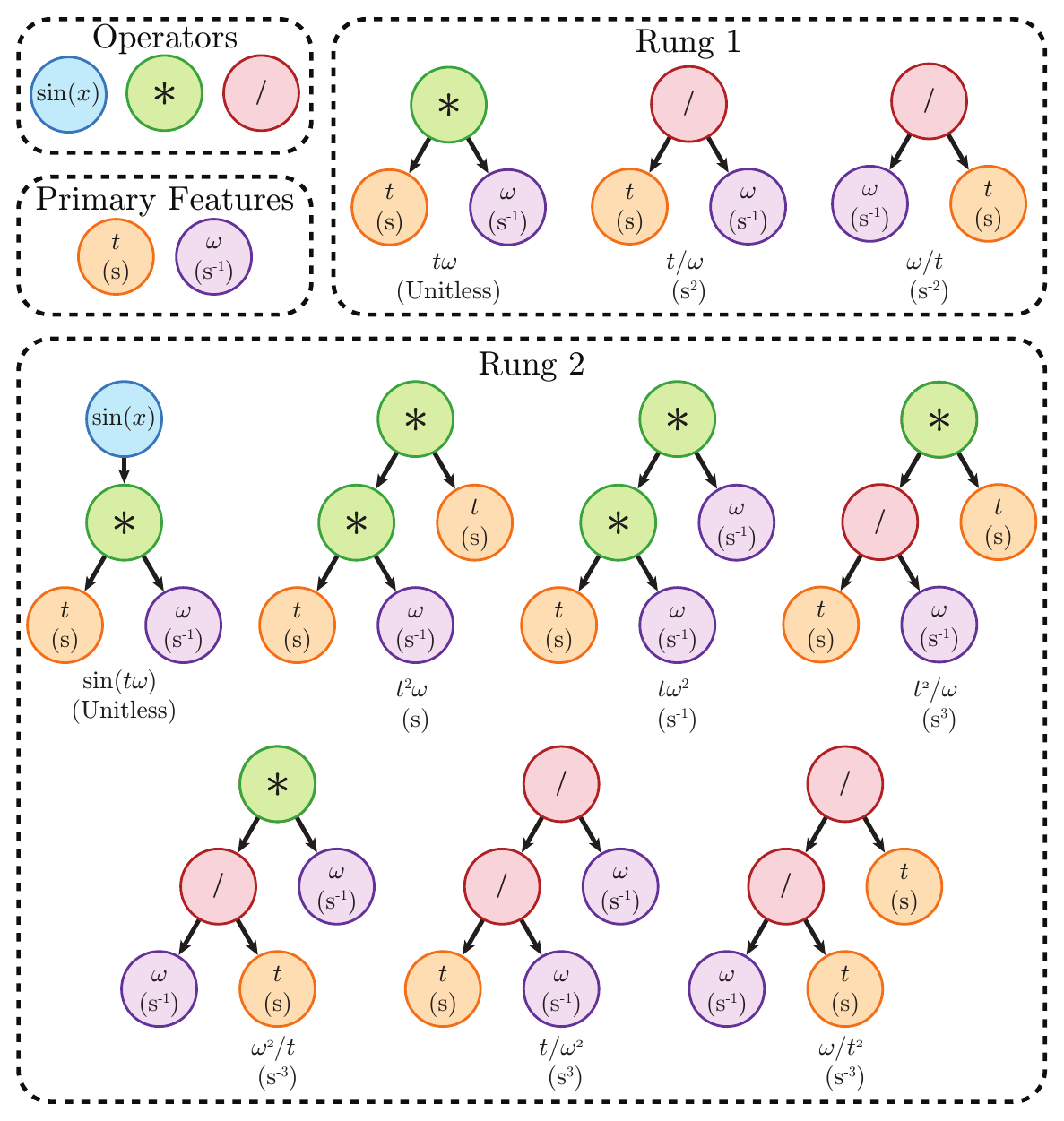}
    \caption{Illustration of how the feature space of SISSO is created. In this example the user selects two primary features $\omega$ (purple) and $t$ (orange) and three operators sin (blue), multiplication (green), and division (red). SISSO then builds up a more complicated expression space by applying the operations onto the existing features by increasing the height of the binary expression trees. Throughout this process the units and ranges of each of the operations are respected.}
    \label{fig:feature_creation_step}
\end{figure*}
Each optimization follows a two or three step process outlined here.
First a local optimization is performed to find the local minimum associated with the initial parameters.
Once at a local minimum an optional global optimization is performed to find any minima that are better than the one initially found.
For these first two steps the parameters are optimized to a relative tolerance of $10^{-3}$ and $10^{-2}$ respectively, with a maximum of five thousand function evaluations.
Finally, a more accurate local optimization is done to a relative tolerance of $10^{-6}$ to find the best parameter set.
For this final optimization, ten thousand function evaluations are allowed.
Additionally, for both the initial and global optimization steps, the parameters are bounded to be in a range between -100 and 100 to improve the efficiency of the optimization, but this restriction is removed for the final optimization.
For all local optimizations, the subplex algorithm~\cite{SUBPLEX}, a faster and more robust variant of the Nelder-Mead Simplex method~\cite{nelder1965simplex}, is used.
The Improved Stochastic Ranking Evolution Strategy algorithm~\cite{IRES} is used for all global optimizations.
Once optimized, only the internal $\alpha$ and $\beta$ parameters are stored in $\hat{\mathcal{P}}$.

Figure~\ref{fig:param_results} illustrates the power of the new parameterization scheme.
For both toy problems represented by analytic Lorentzian and sin functions with some white noise, the non-parametric version of SISSO can not find accurate models for the equations as it can not address the non-linearities properly.
By using this new parameterization scheme, SISSO is now able to accurately find the models as shown in Figure~\ref{fig:param_results}c and d.
However, it is important to note that the more powerful featurization comes at the cost of a significantly increased time to generate the feature space, as the parameterization becomes the bottleneck for the calculations.
Additionally, there can be cases where the parameterization scheme does not find an optimal solution because of too much noise or the optimal parameters being too far away from the initial guesses, as shown in Figure~\ref{fig:param_results}e and f.

\subsection{Building the Complete Feature Set}
With all of the new aspects of the feature representation in place, SISSO++ has a fully parallelized feature set construction that uses a combination of threads and MPI ranks, for efficient feature set construction. 
The basic process of creating new features is illustrated in Figure~\ref{fig:feature_creation_step}, where each new rung builds on top of existing features by adding a new operation on top of existing binary expression trees for the previous rung.
Throughout this process all checks are done to ensure that the units are correct and the domains for each new operation are respected.
Additionally, the code checks for invalid values, e.g., \textsc{NaN} or \textsc{inf}, and some basic simplifications for all features, e.g., features like $\frac{t\omega}{\omega}$ are rejected.
The operators are separated into parameterized and non-parameterized versions of each other to allow for the optional use of the parametric SISSO concepts.

\FloatBarrier

\section{Descriptor Identification}

\subsection{Linear Programming Implementation for Classification Problems}
\label{subsec:di_lp_class}
One of the largest updates to the SISSO methodology is the new, generalized approach for solving classification problems. 
In previous implementations, when finding a classification scheme, SISSO would explicitly build the convex hull, and then calculate the number of points inside the overlap region between different classes and either the normalized overlap volume or separation distance to find the optimal solution.
While this works for two dimensions, finding the overlap volume or separation distance becomes intractable for three or more dimensions, and even defining the convex hull becomes intractable for four or more dimensional classification.
SISSO++ replaces these conditions with an algorithm that determines the number of points inside the convex hull overlap region using linear programming, and explicitly creates a model using linear SVM.
The linear programming algorithm checks for the feasibility of
\begin{align}
&\min 0\\
&\mathrm { s.t. } \sum_{i \in I}^{\min } \alpha_i x_i=x_j, \quad \sum_{i \in I} \alpha_i=1, \quad \alpha_i \geq 0, \ \forall i \in I \mathrm {, }
\end{align}
where $x_i$ is the $i^{th}$ point inside the set of all points of a class $I$, $\alpha_i$ is the coefficient for $x_i$, and $x_j$ is the point to check if it is inside the convex hull.
The above problem is only feasible if and only if $x_j$ lies inside the set of points, $I$, representing a class in the problem.
Here we are optimizing a zero function because the actual solution to this optimization does not matter, only that such a solution can be found, i.e. the constraints can be fulfilled, is important.
The feasibility and linear programming problem is defined using the Coin-CLP library~\cite{Coin-CLP}.
Once the number of points in the overlap region is determined, a linear SVM model is calculated for the best candidates, and used as the new tie-breaking procedure.
The first tiebreaker is the number of misclassified points by the SVM model and the second one is the margin distance.
The SVM model is calculated using libsvm~\cite{libsvm}.

\begin{figure}
    \centering
    \includegraphics{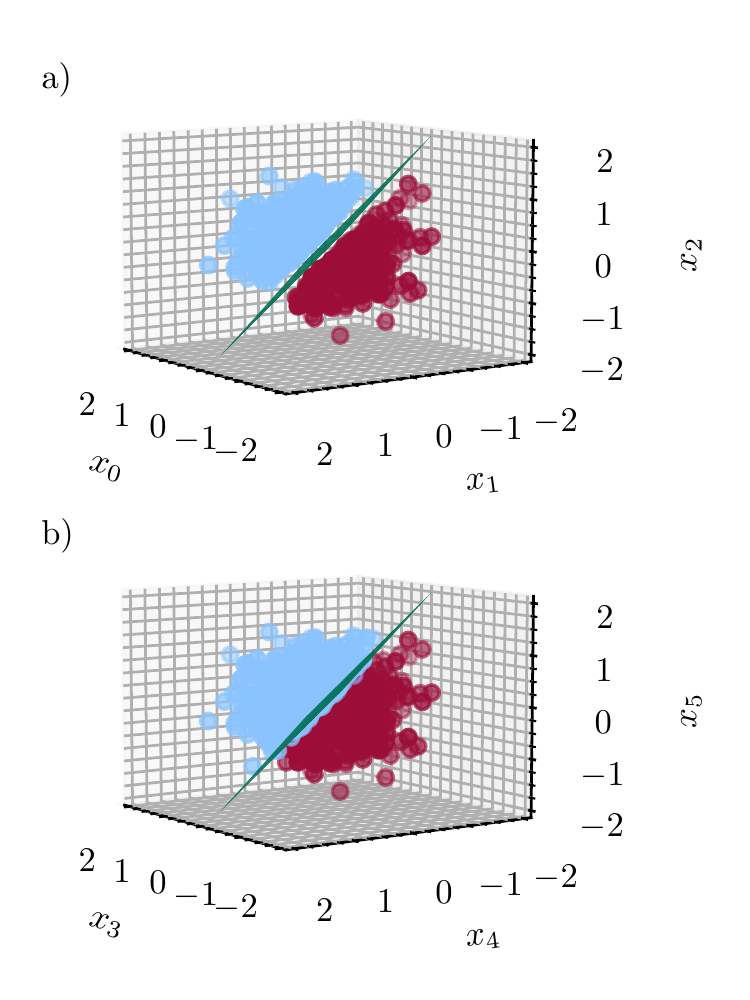}
    \caption{A demonstration of the classification algorithm. A set of 1\smallskip000 points of three features, $x_0$, $x_1$, and $x_2$, are sampled from a Gaussian distribution with a standard deviation of 0.5 and centered at the origin. The set is separated into 2 classes (the red and light blue circles) by the plane $x_0 + x_1 + x_2 = 0$ (green surface). a) All points where $\left|x_1 + x_2 + x_3\right| < 0.4$ are moved to a random point from that has a distance of $0.2 \sqrt{2}$ to $0.35 \sqrt{2}$ from the plane within each of their own classes, and b) the original data stored as $x_3$, $x_4$, and $x_5$. The updated classification algorithm correctly determines that $x_0$, $x_1$, and $x_2$ is the superior classifier, while for the original definitions of only the convex overlap region for three and more dimensions they would be considered equally good. The projections are shown with an elevation angle of 7.5 degrees and an azimuthal angle of 145 deg. }
    \label{fig:class_demo}
\end{figure}

Figure~\ref{fig:class_demo} demonstrates the capabilities of the new classification algorithm.
In this problem, we randomly sample a Gaussian distribution with a standard deviation of 0.5 and centered at the origin to generate one thousand samples for three features, $x_0$, $x_1$, and $x_2$.
These three features and the copied and relabeled to $x_3$, $x_4$, and $x_5$
The samples are then separated into two classes based on if $\left(x_1 + x_2 + x_3\right)$ is greater than (light blue) or less than (red) zero.
For $x_0$, $x_1$, and $x_2$ we move all points where $\left|x_1 + x_2 + x_3\right| < 0.4$ to a random point further away from the dividing plane within each of their classes to ensure a noticeable margin of separation.
For simplicity, we use only the six primary features shown on the axes of Figure~\ref{fig:class_demo}a and b with no generated expressions; however, the rung 2 feature of $\left(x_0 + x_1\right) + x_2$ would be able to completely separate the classes in one dimension.
Using the updated algorithm SISSO can now easily identify that the set $\left\{x_0, x_1, x_2\right\}$ is the better classifier than the set $\left\{x_3, x_4, x_5\right\}$ as the tiebreakers are defined for all dimensions and not just the first two.
More importantly, SISSO can now provide the $n$-dimensional dividing plane found by linear-SVM creating an actual classifier automatically here shown by the green plane.
It is important to note that the same procedure can be done to an arbitrary dimension, but visualization becomes impractical above three.

\subsection{Multiple Residuals}
\label{subsec:mult_res}
\begin{figure*}
    \centering
    \includegraphics{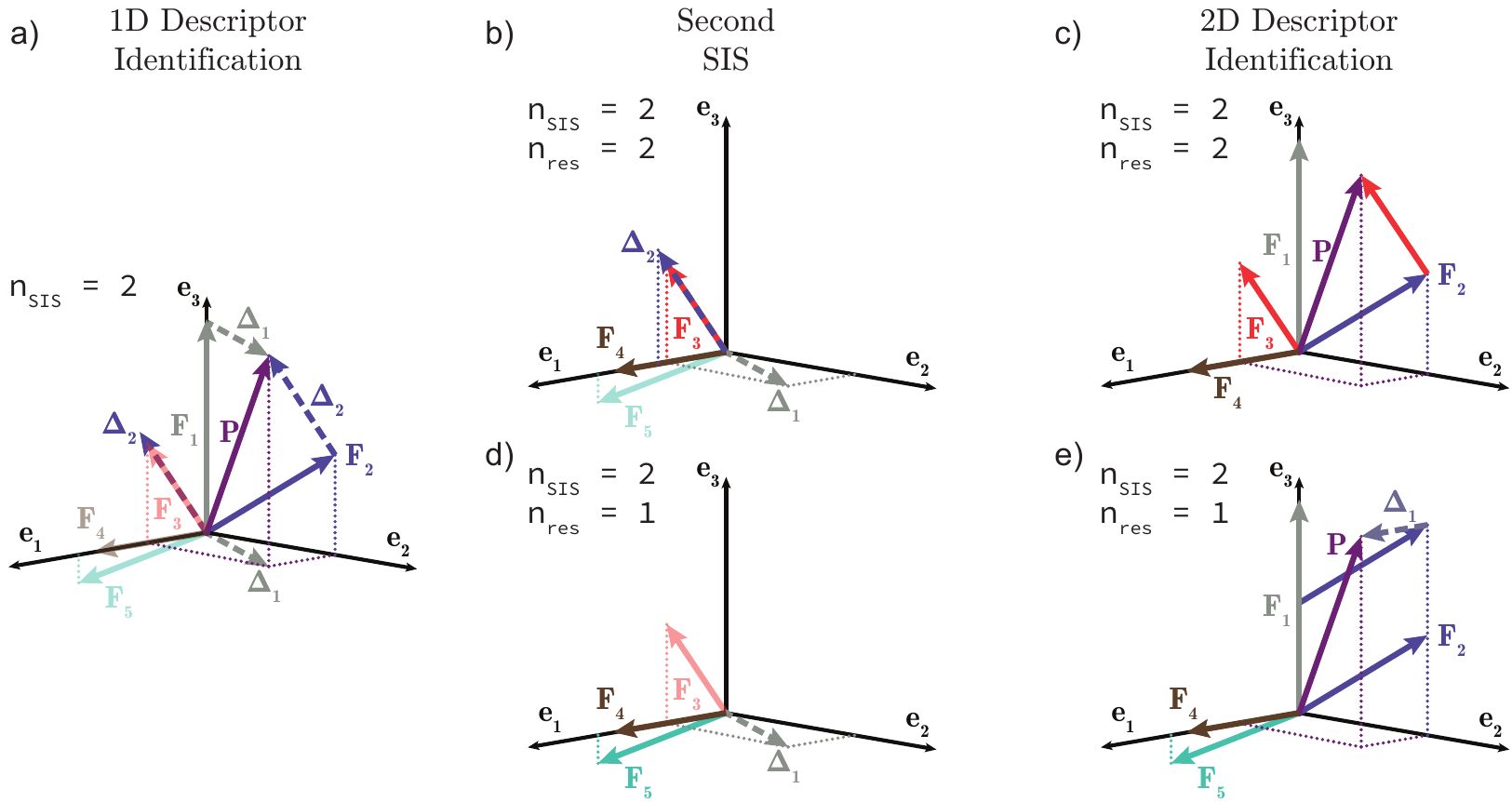}
    \caption{An illustration of how tracking multiple residuals can improve the performance of SISSO. A three dimensional problem space (three training samples) defined by $\mathbf{e_1}$, $\mathbf{e_2}$, and $\mathbf{e_3}$ with five feature vectors $\mathbf{F_1}$ (gray), $\mathbf{F_2}$ (blue), $\mathbf{F_3}$ (red), $\mathbf{F_4}$ (brown), and $\mathbf{F_5}$ (turquoise) for a property vector, $\mathbf{P}$ (purple). For this example the size of the SIS subspace, $n_{sis}$, is two. The residuals for the best ($\mathbf{\Delta_1}$, gray) and second best($\mathbf{\Delta_2}$, blue) one-dimensional models are shown as dashed lines. If the number of residuals, $n_{res}$, is one, then only $\mathbf{\Delta_1}$ is used (d, e) and $\mathbf{F_3}$ will not be selected in the second SIS step. This means the best two dimensional model is not found. However, if $n_{res}=2$ (b, c), then $\mathbf{F_3}$ is selected and the best two dimensional model, a combination of $\mathbf{F_2}$ and $\mathbf{F_3}$, can be found in the second $\ell_0$ step. 
    }
    \label{fig:multires}
\end{figure*}

The second advancement to the descriptor-identification step of the SISSO algorithm is the introduction of a \textit{multiple-residuals} approach to select the expressions (features) for models with a dimension higher than one.
In the original SISSO algorithm~\cite{Ouyuang-2018}, the residual of the previously found model, $\vec{\Delta}_{D-1}^0$, i.e., the difference between the vector storing the values of the property for each sample, $\vec{P}$,  and the estimates predicted by the ($D\! - \! 1$)-dimensional model ($\vec{\Delta}_{D-1}^0 = \vec{P}_{D-1} - \vec{P}$),  is used to calculate the projection score of the candidate features during the SIS step for the best $D$-dimensional model, $s_j^0 = R^2\left(\vec{\Delta}_{D-1}^0, \vec{d}_j\right)$. 
Here, $R$ is the Pearson correlation coefficient, representing a regression problem, and $j$ corresponds to each expression generated during the feature-creation step of SISSO.
In SISSO++\cite{Purcell2022}, we extend the residual definition and use the best $r$ residuals to calculate the projection score: $\mathrm{max}\left(s_j^0, s_j^1, \ldots, s_j^{r-1} \right)$. 
The multiple-residual concept generalizes the descriptor identification step of SISSO, by using information from an ensemble of models to determine which features to add to the selected subspace.
The value of $r$ for a calculation is set as any hyperparmeter via cross-validation

 This process is illustrated in Fig.~\ref{fig:multires} for a three-dimensional problem space (three training samples) defined by $\mathbf{e_1}$, $\mathbf{e_2}$, and $\mathbf{e_3}$ with five candidate features $\mathbf{F_1},\cdots,\mathbf{F_5}$ to describe the property $\mathbf{P}$. 
 In principle, there can be an arbitrary number of feature vectors, but for clarity we only show five. 
 If only a single residual is used, then the selected two dimensional model will comprise of $\mathbf{F_1}$ and $\mathbf{F_4}$, as $\mathbf{F_3}$ is never selected because it has the smallest projection score from the residual of the model found using $F_1$. 
 However, because $\mathbf{F_2}$ has a component along $\mathbf{e_2}$, a better two-dimensional model consisting of a linear combination of $\mathbf{F_2}$ and $\mathbf{F_3}$ exists, despite $\mathbf{F_2}$ not being the most correlated feature to $\mathbf{P}$. 
 When going to higher dimensional feature spaces, it becomes more likely that the feature vectors similarly correlated to the property contain orthogonal information, thus the need for using multiple residuals in SISSO.

\begin{figure}
    \centering
    \includegraphics{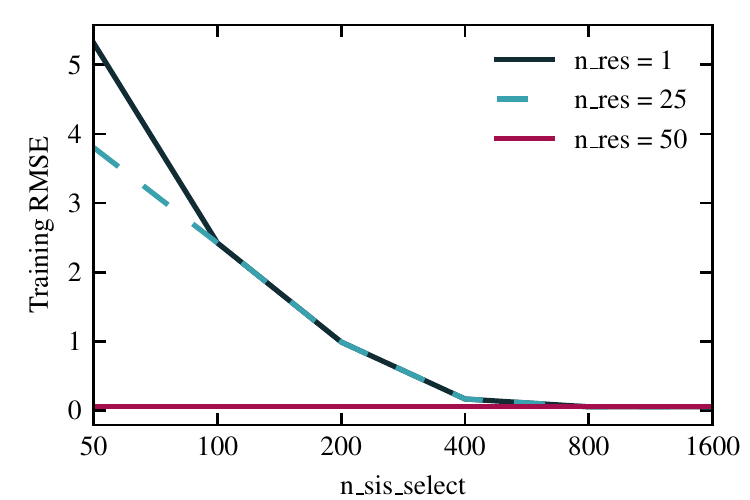}
    \caption{The training RMSE of a two dimensional model against the size of the SIS subspace size for $y=5.5 + 0.4158 d_1 - 0.0974 d_2 + \delta$, where $d_1 = x_0^2 \sqrt[3]{x_1}$, $d_2 = \left|x_2^3\right|$, and $\delta$ is a Gaussian white noise term pulled from a distribution with a standard deviation of 0.05. The solid dark blue line is the error when learning using one residual, the light blue dashed line is the error when learning using twenty five residuals, and the red solid line is the error when learning with fifty residuals.
    }
    \label{fig:mr_conv}
\end{figure}

In order to demonstrate the effect of learning over multiple residuals and get an estimate of the optimal number of residuals and size of the SIS subspace ($n_\text{sis}$), we plot the training RMSE for the two dimensional models for the function$y=5.5 + 0.4158 d_1 - 0.0974 d_2 + \delta$, where $d_1 = x_0^2 \sqrt[3]{x_1}$, $d_2 = \left|x_2^3\right|$, and $\delta$ is a Gaussian white noise term pulled from a distribution with a standard deviation of 0.05 in Fig.~\ref{fig:mr_conv}.
For this problem the best one dimensional descriptor is $\frac{1}{\sqrt[3]{x_3}}$, with $x_3$ being explicitly set to $\left(y + \Delta\right)^{-3}$ and $\Delta$ is Gaussian white noise with a standard deviation of 20.0.
This primary feature was explicitly added in order to ensure that the best one-dimensional model would not be either $d_1$ or $d_2$ in this synthetic problem.
Because of this, when using a single residual the SIS subspace size has to be increased to over 400, before $y$ can be reproduced by SISSO.
However, by increasing the number of residuals to 50, SISSO can now find which features are most correlated to the residual of $d_1$ and it immediately finds $y$.
In a recent paper we published, we further demonstrate this approaches effectiveness for learning models of the bulk modulus of cubic perovskites,~\cite{Foppa2022}.

\section{Conclusions}
In this paper, we described recently developed improvements to the SISSO method and their implementation in SISSO++ code, both in terms of their mathematical and computational details, which constitute a large leap forwards in terms of the expressivity of the SISSO method.
Utilizing these features provides greater flexibility and control over the expressions found by SISSO, and acts as a start to introducing ``grammatical'' rules into SISSO and symbolic regression.
In particular, concepts such as the units and ranges of the formula could be extended to prune the search space of possible expressions for the final models.
We have also described the implementation of {\em Parameteric SISSO}, which considerably opens up the range of possible expressions found by SISSO.
Finally, we discussed two improvements related to the SISSO solver, i.e., a linear programming implementation for the classification problems and the multiple-residuals technique, both providing extended flexibility in the descriptors and models found by SISSO.

\section{Acknowledgements}
T.A.R.P. thanks Christian Carbogno for valuable discussions related to the parametric SISSO scheme and proof reading those parts of the manuscript. T.P. thanks Lucas Foppa for discussions related to the multi-residual approach and proof reading those parts of the manuscript.
This work was funded by the NOMAD Center of Excellence (European Union's Horizon 2020 research and innovation program, grant agreement Nº 951786), the ERC Advanced Grant TEC1p (European Research Council, grant agreement Nº 740233), BigMax (the Max Planck Society's Research Network on Big-Data-Driven Materials-Science), and the project FAIRmat (FAIR Data Infrastructure for Condensed-Matter Physics and the Chemical Physics of Solids, German Research Foundation, project Nº 460197019). T.P. would like to thank the Alexander von Humboldt (AvH) Foundation for their support through the AvH Postdoctoral Fellowship Program.

\section{Conflict of Interest Statement}
The authors have no conflicts to disclose.

\section{Author Contributions}
T.A.R.P. implemented all methods and performed all calculations. T.A.R.P. ideated all methods with assistance from LMG. MS and LMG supervised the project. All authors wrote the manuscript.

\section{Data Availability Statement}
The data that support the findings of this study and all scripts used to generate the figures are openly available in FigShare at \href{https://dx.doi.org/10.6084/m9.figshare.22691857}{DOI: 10.6084/m9.figshare.22691857}.

\bibliographystyle{aip}
\bibliography{main}
\end{document}